\documentclass[journal]{IEEEtran}
\usepackage{multirow,epsfig,fbox,amsfonts,amsmath,multicol,enumitem,algorithm,algorithmic,pifont,graphicx,bm}
\usepackage{booktabs}
\usepackage[caption=false,font=footnotesize]{subfig}
\graphicspath{{Figures}}

\newcommand{\jl}[1]{\textcolor{black}{#1}}

\usepackage{subfig,subfloat}
\usepackage{graphicx}
\usepackage[subfigure]{tocloft}
\usepackage[numbers,sort&compress,comma]{natbib}
\usepackage{hyperref}
\hypersetup{
	colorlinks=true,
	linkcolor=blue
}

\begin{document}
\title{Adaptive Semantic-Enhanced Denoising Diffusion Probabilistic Model for Remote Sensing Image Super-Resolution}
\author{Jialu Sui,
	Xianping Ma,~\IEEEmembership{Student Member,~IEEE,}
    Xiaokang~Zhang,~\IEEEmembership{Member,~IEEE,}
    Man-On~Pun,~\IEEEmembership{Senior~Member,~IEEE}
    and~Hao Wu
	\thanks{This work was supported in part by the National Key Research and Development Program of China under Grant 2022YFB3903502 and 2018YFB1800800, the Basic Research Project under Grant HZQB-KCZYZ-2021067 of Hetao Shenzhen-HK S$\&$T Cooperation Zone, Shenzhen Outstanding Talents Training Fund 202002, Guangdong Research Projects under Grant 2017ZT07X152 and 2019CX01X104, the Guangdong Provincial Key Laboratory of Future Networks of Intelligence under Grant 2022B1212010001, the National Natural Science Foundation of China under Grant 42371374 and 41801323. \textit{(Corresponding authors: Man-On Pun; Xiaokang Zhang)}}
	\thanks{Jialu Sui, Xianping Ma and Man-On Pun are with the School of Science and Engineering, The Chinese University of Hong Kong, Shenzhen, Shenzhen 518172, China (e-mail: jialusui@link.cuhk.edu.cn; xianpingma@link.cuhk.edu.cn; SimonPun@cuhk.edu.cn).}
    \thanks{Xiaokang~Zhang is with the School of Information Science and Engineering, Wuhan University of Science and Technology, Wuhan 430081, China (e-mail: natezhangxk@gmail.com).}
    \thanks{Hao~Wu is with the College of Urban and Environment Sciences, Central China Normal University, Wuhan 430079, China (e-mail: haowu@mail.ccnu.edu.cn).}
    }
\maketitle
  
\begin{abstract}
Remote sensing image super-resolution (SR) is a crucial task to restore high-resolution (HR) images from low-resolution (LR) observations. Recently, the Denoising Diffusion Probabilistic Model (DDPM) has shown promising performance in image reconstructions by overcoming problems inherent in generative models, such as over-smoothing and mode collapse. However, the high-frequency details generated by DDPM often suffer from misalignment with HR images due to the model's tendency to overlook long-range semantic contexts. This is attributed to the widely used U-Net decoder in the conditional noise predictor, which tends to overemphasize local information, leading to the generation of noises with significant variances during the prediction process. \jl{To address these issues, an adaptive semantic-enhanced DDPM (ASDDPM) is proposed to enhance the detail-preserving capability of the DDPM by incorporating low-frequency semantic information provided by the Transformer. Specifically, a novel adaptive diffusion Transformer decoder (ADTD) is developed to bridge the semantic gap between the encoder and decoder through regulating the noise prediction with the global contextual relationships and long-range dependencies in the diffusion process.} Additionally, a residual feature fusion strategy establishes information exchange between the two decoders at multiple levels. As a result, the predicted noise generated by our approach closely approximates that of the real noise distribution.
Extensive experiments on two SR and two semantic segmentation datasets confirm the superior performance of the proposed ASDDPM in both SR and the subsequent downstream applications. The source code will be available at \href{https://github.com/littlebeen/ASDDPM-Adaptive-Semantic-Enhanced-DDPM}{https://github.com/littlebeen/ASDDPM}.
\end{abstract}


\IEEEpeerreviewmaketitle

\section{Introduction}\label{sec:intro}
Satellite-based remote sensing technology has a broad range of applications on earth observations \cite{remote1,zhang2023cross}. However, many such remote sensing applications demand very high spatial-resolution images. Unfortunately, the hardware complexity required to acquire such high-resolution images is usually prohibitively expensive. Furthermore, satellite images are often distorted by various impairments due to their ultra-long-range imaging nature, such as atmospheric disturbance and noise from the imaging system, resulting in inaccurate and unreliable information \cite{overview4}. As a result, the spatial resolution of the satellite images has to be enhanced to meet the requirements of practical applications \cite{wu2024df4lcz}. 

Recently, single-image super-resolution (SISR) \cite{overview, overview2} for remote sensing at the sub-pixel level has gained increasing attention \cite{max1, max2}. In lieu of generating high-resolution images with expensive sensing equipment, the SISR techniques are designed to improve the resolution of existing low-resolution (LR) RGB \cite{sui2023gcrdn, dtrn} or hyperspectral \cite{li1,li2} images by exploiting complementary information. Driven by the recent advances in deep learning (DL) technology, DL-based SISR has been intensively studied. Generally speaking, these DL-based SISR methods can be divided into two categories \cite{deep,deep2}, namely discriminative and generative models.

The discriminative models adopt the convolutional neural networks (CNNs) as the baselines in a fully supervised manner to obtain realistic SR images \cite{cnn2, edrn}. For instance, SRCNN \cite{SRCNN} first introduced CNN into SISR by learning an end-to-end mapping between the LR and HR images. Furthermore, deeper networks with skip connections have been shown more effective in learning robust representations. In \cite{cnn4}, a residual aggregation and split attentional fusion network was proposed for high-quality SR of remote sensing images whereas WTCRR was developed in \cite{cnn1} to lessen the training complexity of the deep network. Despite their acceptable performance, these methods are ineffective in capturing global information due to the convolutional counting process, making it challenging to exploit long-range relationships between different ground objects. To tackle this problem, the self-attention mechanism has been adopted into the convolutional operations to generate contextualized features \cite{multilevel}. For instance, NLSN \cite{nlsn} adopted a novel non-local sparse attention with an attention bucket calculated with locality-sensitive hashing embedded into several residual blocks to extract the global context. However, discriminative models trained by distance-based loss functions such as the $L_{1}$ and $L_{2}$ loss functions in the pixel level could be biased, entailing high-frequency information loss. Those image similarity-oriented loss functions of discriminative models have been developed to provide reasonable Peak Signal-to-Noise Ratio (PSNR) and Structural Similarity Index (SSIM) metrics without taking into account the actual requirements from the subsequent applications. As a result, such similarity-oriented loss functions incur loss of detail textures in the reconstructed images, which adversely impacts the subsequent applications.

Recently, generative models, particularly generative adversarial networks (GANs) \cite{gan1,gan2}, have become a popular choice for large-scale SISR to optimize image reconstruction. In remote sensing, EEGAN \cite{EEGAN} focused on edge enhancement by purifying the noise-contaminated components with mask processing. SWCGAN \cite{swcgan} utilized a hybrid network architecture consisting of convolutional and Swin Transformer layers as the generator, while a pure Swin Transformer is adopted as the discriminator. Specifically, the LR images are encoded and then decoded in the generator whereas the generated results are subsequently fed into the discriminator to distinguish whether the results are ``real" enough for human perception with the generative adversarial learning strategy. Therefore, the HR images restored by the GAN-based method are more realistic in human visual perception. However, GAN-based models have encountered challenges such as model collapse, unstable training, and vanishing gradients. Moreover, GAN-based methods rely on pixel-wise distance or similarity-based losses and often yield over-smoothed outcomes, limiting their applications in downstream visual tasks like semantic segmentation.
 
More recently, a novel generative model called the denoising diffusion probabilistic model (DDPM) \cite{ddpm,ddpm2} has shown promising performance in computer vision tasks with excellent overall reconstruction quality in image synthesis. The DDPM consists of a diffusion process and a reverse process. More specifically, the diffusion process injects noise into images using data distribution transfer to the latent variable distribution, while the diffusion process generates HR images through noise projection. DDPM is trained by optimizing a variant of the variational lower bound, which is robust against the mode collapse problem of GANs and mitigates the limitation of the PSNR-based loss function of the discriminative model \cite{ddpm}. \jl{However, the high-frequency details generated by the DDPM-based methods may misalign with the target image, which incurs performance degradation. This is because the U-Net-based conditional noise predictor (CNP) excessively leverages high-frequency image components \cite{wang2020high} related to local geographical features while neglecting global information across the entire image. However, for the generative model, only high-frequency information is apparently insufficient as the real world is made up of high- and low-frequency signals. Despite the loss function optimizes the CNP toward the true distribution, there exists an inherent limitation in the model's structure. As a result, this overuse of local information generates noises with notable variances during the noise prediction process, leading to undesirable textures in the reconstructed images. In particular, for complex remote sensing image generation tasks, global information has more important guiding significance. The main reason is that remote sensing images generally have global similarities, such as texture, color, and shape. Thus, further investigation based on frequency is required to reconstruct high-quality HR images to meet the demands of downstream remote sensing tasks.} 

\jl{In this paper, we propose a straightforward yet effective solution named Adaptive Semantic-enhanced DDPM (ASDDPM) to tackle the previously mentioned issues by employing a dual-decoder strategy in the CNP. Based on our observation that there is inevitably a semantic gap between the encoder and decoder, we consider that the decoder may struggle to reconstruct fine-grained details from the learned features. Therefore, we focus on enhancing the decode to improve the semantic representation of the LR image. Specifically, an adaptive diffusion Transformer decoder (ADTD) is proposed to regulate the generation of detailed information by integrating global contextual relationships and long-range dependencies between homogeneous objects. This enhancement enables the model to produce noise distributions more similar to real-world scenarios during training. In other words, we introduce Multi-head Self-Attentions (MSA)-based modules to generate low-frequency information \cite{park2022vision}.} On this basis, a residual feature fusion approach is developed to enable efficient feature exchange between the two decoders at multiple levels.

semantic representation ability

The main contributions of this work are summarized as follows.
\begin{enumerate}
\item \jl{A novel DDPM is proposed by incorporating a low-complexity ADTD into the CNP to bridge the semantic gap between the encoder and decoder. This architecture incorporates a low-frequency semantic information generation module into the decoding stage, aiming to enhance the DDPM model from a frequency perspective.}

\item The cooperation between the U-Net decoder and the ADTD built into a feature fusion module effectively preserves the semantic information of LR images, ultimately producing SR images with sharpened details and textures. This emphasis on perceptual qualities improves performance for downstream semantic segmentation tasks and tackles the problem of generative models that obtain high SSIM and PSNR values while yielding over-smoothed outcomes.

\end{enumerate}

The rest of the paper is organized as follows. Section~\ref{sec:related Work} provides a brief review of existing DL-based SR models followed by elaboration on the proposed ASDDPM in Section~\ref{sec:method}. After that, the performance of the proposed ASDDPM and other state-of-the-art SR methods is evaluated and discussed in Section~\ref{sec:experiments} before the conclusion is given in Section~\ref{sec:conclusion}.

\section{Related Work}\label{sec:related Work}

\begin{figure*}
  \centering
   \includegraphics[width=0.95\linewidth]{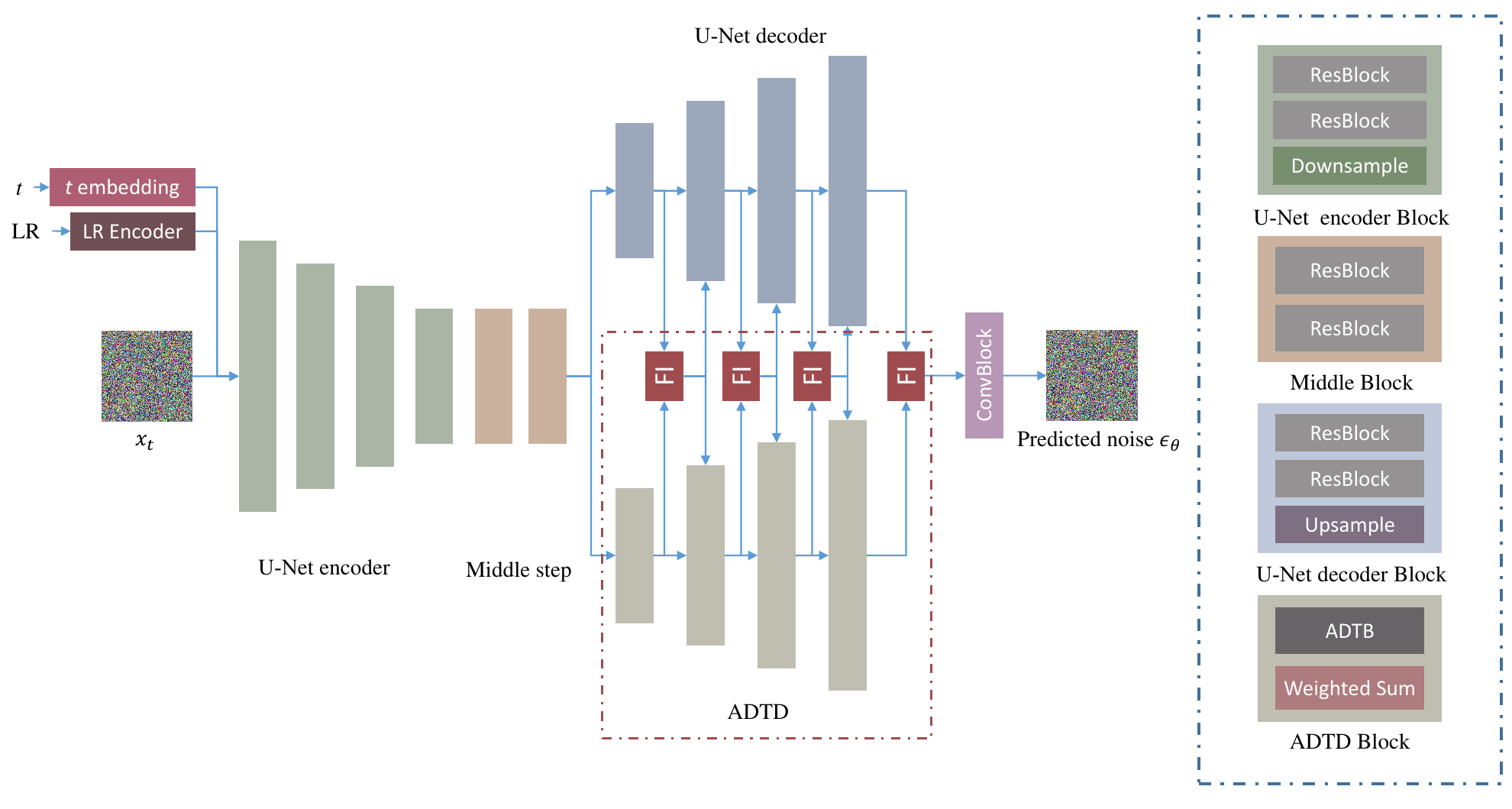}
   \caption{The architecture of the CNP in the proposed ASDDPM. The four module types, U-Net encoder, Middle step, U-Net decoder, and ADTD, are denoted using different colors and further illustrated using the legends at the right part. The main contributions of the proposed ASDDPM are highlighted and represented by the red dotted box. Moreover, ResBlock means the residual block and ConvBlock contains several convolutional layers to make final channel restoration}.
   \label{fig:ALL}
\end{figure*}

\subsection{Discriminative Models}
Discriminative models have been widely used for enhancing image perceptual quality for several decades. The success of AlexNet and VGG-net in image recognition tasks has motivated the development of many CNN-based methods \cite{vdsr} to address SISR problems. Recently, deeper networks with residual dense learning strategy \cite{EDSR} have attracted intensive research attention for their capability of producing explicit content. For instance, RDN \cite{rdn} presented a combination of the residual and dense blocks to generate HR images with precise details by adaptively learning features. In the field of remote sensing, GEDRN \cite{GEDRN} employed share-source residual structure and non-local operations to learn low-frequency information. However, since CNN exploits and aggregates enriched local features using the local receptive fields in the convolutional layer, it is ineffective in extracting global features. To cope with this shortcoming, the Transformer architecture was proposed to fully exploit long-distance correlation. For instance, SwinIR \cite{swinir} applied several Transformer blocks with a residual connection as an encoder to enable global feature extraction. Recently, hybrid Transformer-CNN models have been developed for remote sensing SR problems. For instance, Interactformer \cite{interactformer} adaptively learned both global and local features by exploiting two branches of encoders, including one CNN branch and one Transformer branch. Furthermore, \citet{TranSMS} developed a dual branch network called TranSMS by capitalizing on a vision transformer module and a dense convolutional module to capture contextual relationships in LR images and localize SR image features, respectively. However, the complex encoding structures of these hybrid discriminative models usually require tedious designs of loss functions and training strategies for model optimization. As a result, they tend to generate SR images with poor perceptual quality due to their weak image reconstruction ability \cite{wang2022comprehensive}.

\subsection{Generative Models}
Motivated by the discussions above, generative models have been developed as an alternative solution by minimizing perceptual differences between the ground truth and output images. These generative models are designed to generate images of high visual quality, i.e. more visually appealing to human observers. Many generative models, such as GAN and DDPM, were introduced into SISR for image reconstruction. For GAN-based methods \cite{SRGAN, gan3}, a perceptual loss function consisting of an adversarial loss and a content loss is employed to guide the generator network training process. More specifically, the adversarial loss is generated by a convolutional discriminator network trained to differentiate between the SR and HR images. Endowed with the generative adversarial learning strategy, \cite{rrsgan} devised a reference-based remote sensing GAN model called RRSGAN to achieve better quantitative and visual results over various scenarios. More recently, MAGAN \cite{MAGAN} was developed for remote sensing images by exploiting multi-attention operation. Despite the fact that the adversarial loss can generally improve the perceptual quality of resulting images, it is computed based on the content loss at the pixel level. As a result, GAN-based methods still suffer from over-smoothing, instability, and model collapse problems \cite{wang2022comprehensive}. 

The DDPM-based SISR models have recently been developed to produce detailed texture information in HR images. For instance, SR3 \cite{SR3} employed the LR images to predict image noise, while SRDiff \cite{srdiff} replaced the LR images with feature maps extracted by a CNN-based LR encoder using a residual-in-residual dense block (RRDB) \cite{ESRGAN}. \jl{Residual diffusion \cite{DocDiff} predicted the residual information, including high-frequency information and edges between the ground truth and the LR images.} Furtheremore, \cite{latentdiff} applies the diffusion process in the latent space to reduce the computational resources while retaining the reconstruction quality and flexibility. \jl{However, these DDPM-based models may produce inaccurate details due to the overuse of high-frequency information related to local features while generating noises with notable noise variances by the U-Net-based CNP. Based on the residual diffusion framework, our model solves the problem through introducing ADTD and a residual feature fusion into the CNP, which incorporates the global contextual relationships to bridge the gap between the encoder and decoder. With the enhanced decoder, our model could generate detailed information while maintaining image accuracy.}

In summary, current SISR methods focus on enhancing the pixel-wise similarity between the resulting images and ground truths. As a result, despite the fact that the resulting images possess impressive PSNR and SSIM values, they usually suffer from over-smoothing problems and fail to meet the image quality required by downstream applications such as semantic segmentation. 

\section{Methodology}\label{sec:method}

\subsection{ASDDPM Architecture}

Based on DDPM, the proposed ASDDPM consists of two Markov chains, including one forward chain for the diffusion steps and one reverse chain for the reverse steps, respectively. The forward chain is designed to transform the data distribution into a latent variable distribution, such as the Gaussian distribution, whereas the reverse chain works in the opposite direction.

\begin{figure*}
	\centering
	\includegraphics[width=0.8\linewidth]{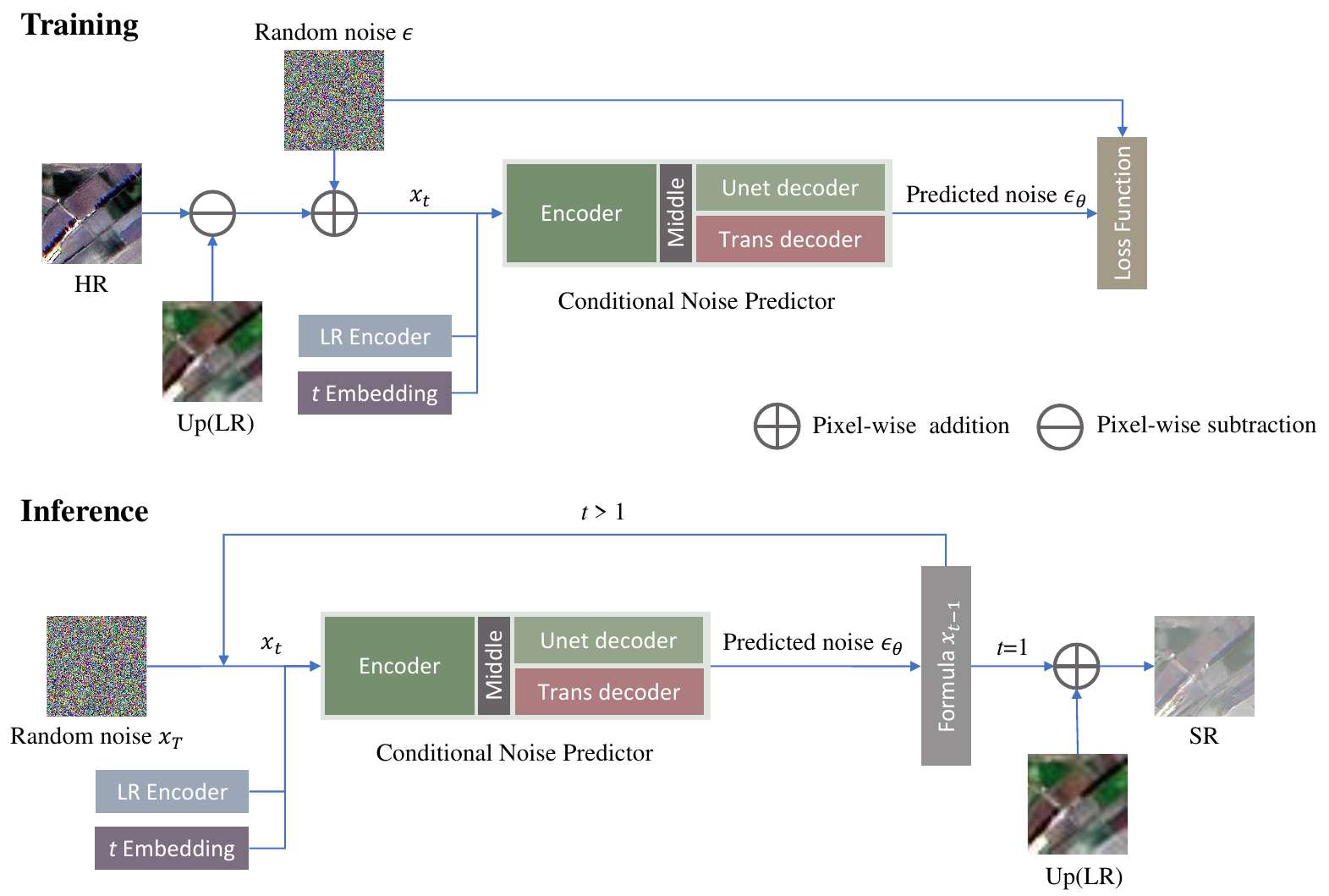}
	\caption{The process of Inference and Training in our ASDDPM. Trans decoder means ADTD.}
	\label{fig:diffusion}
\end{figure*}

\subsubsection{Diffusion process}

Given a data distribution $x_0 \sim q(x_0)$, the diffusion process generates a series of $x_1,x_2,\dots,x_T$, conditioned on $x_0$ where $T$ is the number of diffusion steps. The process could be formulated as follows:
\begin{equation}
  q(x_1,\dots,x_T|x_0) = \prod_{t=1}^{T}q(x_t|x_{t-1}),
\end{equation}
where
\begin{equation}
  q(x_t|x_{t-1}) = {\cal N}(x_t;\sqrt{1-\beta_t}x_{t-1},\beta_t{\bm I}),
\end{equation}
with ${\bm I}$ and ${\cal N}$ being an all-one matrix and the Gaussian distribution, respectively.

According to a variance schedule $\{\beta_1,\dots,\beta_T \}$ with $\beta_t \in \left\{1,0\right\}$ for $t=1,2,\cdots,T$, Gaussian noise $\epsilon \in {\cal N}(0,{\bm I})$ is inserted into the data in the sequence. In other words, each $x_t$ is calculated from $x_{t-1}$. With $a_t=1-\beta_t$ and $\bar{a}_t=\prod_{s=0}^ta_s$, it can be shown that $q(x_t|x_0)$ takes the following form \cite{ddpm,ddpm2}:
\begin{equation}
  q(x_t|x_0) = {\cal N}\left(x_t;\sqrt{\bar{a}_t}x_0,(1-\bar{a}_t){\bm I}\right).
\end{equation}

As a result, every $x_t$ can be directly calculated through $x_0$ using the following formula:
\begin{equation}
  x_t = \sqrt{\bar{a}_t}x_0+\sqrt{1-\bar{a}_t}\epsilon.
\end{equation}

\subsubsection{Reverse Process}
The reverse process converts the data from $p_\theta(x_T)$ to $p_\theta(x_0)$ parameterized by $\theta$. The prior distribution is chosen as $p(x_t)={\cal N}(x_t;0,{\bm I})$ since $q(x_t)$ approximately follows ${\cal N}(x_t;0,{\bm I})$ in the diffusion process. As a result, the learnable transition kernel is given by:
\begin{equation}
  p_\theta(x_0, \dots, x_{T-1}|x_T) = \prod_{t=1}^{T}p_\theta(x_{t-1}|x_t),
\end{equation}
where
\begin{equation}
  p_\theta(x_{t-1}|x_t) = {\cal N}(x_{t-1};\mu_\theta(x_t,t),\sigma_\theta {(x_t,t)}^2{\bm I}),
\end{equation}
with $\theta$ being the model parameters. Furthermore, $\mu_\theta$ and $\sigma_\theta$ are the mean and variance of the Gaussian distribution, respectively.

The model is then trained to minimize the difference between the reverse Markov chain and the reversal of the forward Markov chain in the actual time. The Kullback-Leibler (KL) divergence is commonly employed to calculate the difference of the log-likelihood $p_\theta(x_0,x_1,\dots,x_T)$ and $q(x_0,x_1,\dots,x_T)$ as:

\begin{footnotesize}
\begin{eqnarray}
 &&\mathbb{E}\left[-\log p_\theta(x_0)\right]\nonumber\\&\le& KL\left(p_\theta(x_0,\dots,x_T),q(x_0,\dots,x_T)\right) \\\nonumber
 &=&\mathbb{E}_q\Big[D_{KL}\left(q(x_T|x_0)||p(x_T)\right) + \\
 &&\sum_{t>1}D_{KL}\left(q(x_{t-1}|x_t,x_0)||p_\theta(x_{t-1}|x_t)\right)-\log p_\theta(x_0|x_1)\Big] .
\end{eqnarray}
\end{footnotesize}

For notational simplicity, the loss function denoted as $\ell$ can be simplified as follows after ignoring those items independent of $\theta$:
\begin{equation} \label{eq1}
 \min_{\theta}{\ell_{t-1}} = \mathbb{E}_{x_0,\epsilon,t}\Big[{||\epsilon-\epsilon_\theta(x_t,t)||}^2\Big],
\end{equation}
where $\epsilon_\theta$ is the noise predictor.

Finally, assuming $z \sim {\cal N}(0,{\bm I})$, we can express $x_{t-1}$ as:
\begin{equation}  \label{eq2}
x_{t-1}=\mu_\theta(x_t,t)+\sigma_\theta{(x_t,t)}^2z.
\end{equation}

\subsection{Conditional noise predictor (CNP)}

The CNP is designed to predict noise $\epsilon_\theta$ added at each diffusion time step by exploiting $x_t$. As illustrated in Fig.~\ref{fig:ALL}, the CNP consists of three main components: an encoder, a middle module, and a dual decoder. Initially, $x_t$, the output of a pre-trained LR encoder and the diffusion $t$ embedding obtained from the diffusion time step through positional encoding \cite{attention} are fed into a U-Net encoder. The LR encoder is a pre-trained SR model designed to improve training stability and convergence behaviors by providing coarse features. Next, the features derived from the U-Net decoder are passed through the middle step that comprises two residual blocks. Finally, the output from the middle step is fed into a proposed dual decoder composed of a U-Net decoder and an ADTD. More specifically, the two decoders are connected through the feature integration (FI) modules \cite{ESanet}, facilitating residual feature fusion in each layer. As a result, the model can learn to focus on important features of selected modality while adaptively suppressing less important features by exploiting the FI modules. In summary, the proposed CNP can effectively process $x_t$ while predicting the corresponding noise by leveraging both the \jl{high- and low-frequency information form} U-Net and Transformer architectures, which results in improved SR image reconstruction.

\subsubsection{U-Net Encoder}
The input to the encoder is first processed by a convolution head comprising a convolutional layer and a Mish activation function to increase the number of channels. The features are then passed through the contracting step consisting of four blocks designed to improve the channel dimensionality. Each block comprises two residual blocks and a downsampling layer. The features generated by the LR Encoder are fused with the hidden state produced by the first two residual blocks in the first block of the contracting step. In addition, the output from each block in the contracting step is stored and added to the corresponding output of the expansive step in the decoder.

\subsubsection{The proposed dual-decoder Structure}
The U-Net decoder, as the expansive step, consists of blocks comprising two residual blocks and an upsampling layer to decrease the channel dimensionality. The $t$ embedding guides the feature computation in each residual block in the whole process. \jl{It provides diverse high-frequency image components for local information reconstruction. Furthermore, the proposed ADTD is designed to enhance the feature expression with low-frequency semantic information by exploiting the scaling properties inherent in the transformer architecture.} More specifically, the ADTD comprises a series of Adaptive Diffusion Transformer Block (ADTB) and weighted sum modules. In particular, the weighted sum modules are designed to incorporate the output from the corresponding layer in the encoder, thereby expanding the channel size and facilitating feature fusion. This approach allows the model to capture more informative representations by leveraging the features learned in the encoder, leading to improved performance.

\begin{figure}[h]
	\centering
	\includegraphics[width=\linewidth]{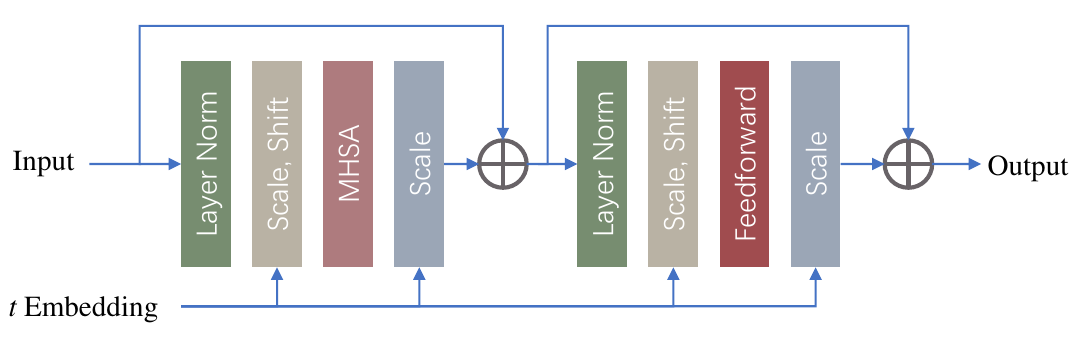}
	\caption{The architecture of ADTB.}
	\label{fig:Dit}
\end{figure}

As shown in Fig.~\ref{fig:Dit}, the proposed ADTB contains two modules of similar structures. More specifically, the first module performs layer normalization, scaling, and shifting operations \cite{scale} followed by a multi-head self-attention (MHSA) and a scaling operation. The MHSA module enables the model to learn long-range dependencies and effectively integrate information from different spatial locations. The second module in ADTB is similar to the first module, except that the MHSA module is replaced by a point-wise feedforward layer designed to learn more complex relationships between the input features. In particular, the proposed ADTB derives dimension-wise scale and shift parameters from the sum of the time step embedding. These parameters will be used to control the noise power.

\subsubsection{Residual Feature Fusion}

\begin{figure}[h]
  \centering
   \includegraphics[width=1\linewidth]{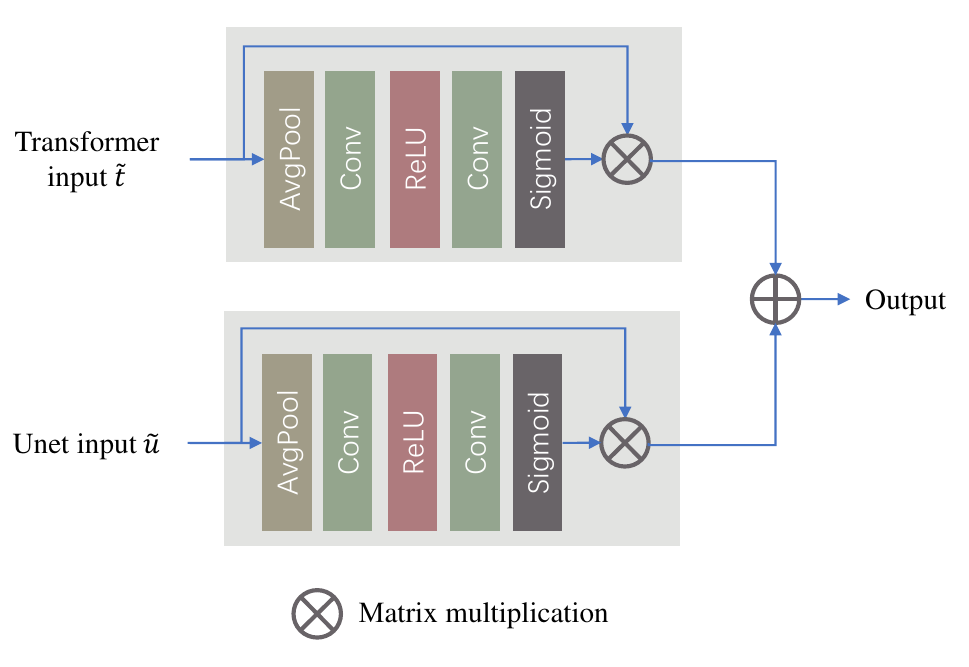}
   \caption{The architecture of FI module.}
   \label{fig:FI}
\end{figure}

The U-Net decoder and the ADTD are connected through a novel residual feature fusion strategy to fuse extracted features, \jl{which efficiently integrates the high- and low-frequency information from each layer of the two decoders.} As shown in Fig.~\ref{fig:FI}, the output from each layer of the ADTD is fed into the FI module and processed by a series of operations, including AvgPool, convolution layers, ReLU activation, and Sigmoid activation. Similarly, the output from each layer of the U-Net decoder is processed in the same fashion in the FI module. Finally, the resulting outputs are fused together through an addition operation as the final FI module output. Mathematically, the operation of the FI module can be expressed as:

\begin{equation}
  {t}_i = \tilde{t}_{i-1}+FI(\tilde{t}_{i-1},\tilde{u}_{i-1}),
\end{equation}
\begin{equation}
  {u}_i = \tilde{u}_{i-1}+FI(\tilde{t}_{i-1},\tilde{u}_{i-1}),
\end{equation}
where $\tilde{t}_i$ and $\tilde{u}_i$ represent the output from the $i$-th layer in the ADTD and the U-Net decoder, respectively. Furthermore, ${t_i}$ and ${u_i}$ denote the input to the $i$-th layer of the two decoder. FI is the feature integration module.

\subsection{Training and Inference}
As illustrated in Fig.~\ref{fig:diffusion}, the training process starts with computing the difference between the upsampled LR images and HR images before a random noise $\epsilon$ is added to the difference. After that, the resulting sum denoted as $x_t$ is fed into the CNP together with $t$ and LR features extracted by the LR encoder \cite{ESRGAN}. More specifically, $\epsilon$ and $t$ are sampled from the standard Gaussian distribution and the integer set $\{1,\dots, T\}$, respectively. Finally, the CNP is optimized by minimizing the loss function shown in  Eq.~(\ref{eq1}).

For the inference process, Gaussian-distributed random noise denoted as $x_T$ is first generated and fed into the reverse process where $T$ is the number of diffusion steps. In each time step, the LR encoder extracts the features from LR images and feeds these features into the CNP together with $x_t$. After the CNP, the predicted noise is utilized to generate $x_{t-1}$ with $x_t$ for $t>0$ using Eq.~(\ref{eq2}). Finally, $x_0$ and upsampled LR images are added together to derive the SR images.

\section{Experiments}\label{sec:experiments}
\subsection{Datasets and Metrics}
In this section, extensive experiments are performed to confirm the effectiveness of the proposed model using four satellite remote sensing datasets, namely OLI2MSI \cite{OLI2MSI}, Alsat \cite{Alsat}, Vaihingen and Potsdam \cite{Potsdam}. As shown in Table \ref{tab:dataset}, OLI2MSI is composed of $5225$ and $100$ pairs of Landsat8-OLI ($10$~m resolution) and Sentinel2-MSI ($30$~m resolution) images for training and testing, respectively. For the Alsat, there are $2182$ pairs of HR images of $2.5$~m resolution and $256\times 256$ and LR images of $10$~m resolution and $64\times 64$ in the training set, and a total of 577 pairs in three test sets, namely ``agriculture", ``urban", and ``special". Vaihingen and Potsdam are two well-known segmentation datasets commonly used for performance evaluation on semantic segmentation models. The segmentation labels comprise six classes, namely ``roads", ``buildings", ``cars", ``low vegetation", ``trees", and ``clutter". 

\begin{table}[h]
	\centering
	\caption{The detailed information of the experimental SR datasets. }
	\setlength{\tabcolsep}{2mm}{
		\begin{tabular}{cccccc}
			\hline
			\textbf{Name}&\textbf{training pairs}&\textbf{testing pairs}&\textbf{scale factor} &\textbf{resolution}\cr
			\hline
                OLI2MSI& 5225 & 100 & 3 & 10 m \\
                Alsat & 2182 & 577 & 4 & 2.5 m  \\
                Vaihingen & 1599 & 1251 & 4 &0.09 m  \\
                Potsdam & 4418 & 2209 & 4 & 0.05 m  \\
			\hline
	\end{tabular}}
	\label{tab:dataset}
\end{table}

\begin{table*}
	\centering
	\caption{The quantitative experimental results on the OLI2MSI and Alsat datasets. The best and second-best are bolded and underlined, respectively.}
	\setlength{\tabcolsep}{4mm}{
		\begin{tabular}{c|cccc|cccc}
			\hline
			\multirow{2}{1cm}{\textbf{Method}}&
			\multicolumn{4}{c|}{\textbf{OLI2MSI}}&\multicolumn{4}{c}{\textbf{Alsat}}\cr
			&\textbf{PSNR}$\uparrow$&\textbf{SSIM}$\uparrow$&\textbf{FID}$\downarrow$&\textbf{LPIPS}$\downarrow$&\textbf{PSNR}$\uparrow$&\textbf{SSIM}$\uparrow$&\textbf{FID}$\downarrow$&\textbf{LPIPS}$\downarrow$\cr
			\hline
                RDN     & \textbf{35.165} & \textbf{0.9049} & 380.210 & 0.0456 & \textbf{15.939} & \underline{0.2681} & 6726.059 & 0.4149   \\
			NLSN    & \underline{35.142} & \underline{0.9035} & 402.998 & 0.0474 & 15.862 & 0.2668 & 6924.580 & 0.4217   \\
                TranSMS & 34.944 & 0.8993 & 428.363 & 0.0505 & \underline{15.929} & 0.2673 & 6418.498 & 0.4021  \\
			SRGAN   & 34.727 & 0.8979 & 413.628 & 0.0495 & 15.879 & 0.2658 & 6769.798 & 0.3968  \\
                Beby-GAN& 34.507 & 0.8898 & 186.267 & 0.0381 & 15.848 & 0.2677 & 5297.296 & 0.3951  \\
			ESRGAN  & 34.295 & 0.8849 & 194.771 & 0.0293 & 15.839 & \textbf{0.2683} & 5949.684 & 0.3464  \\
                Dit     & 33.622 & 0.8675 & 644.122 & 0.0644 & 14.509 & 0.2510 & 6771.076 & 0.3644  \\
                EDiffSR & 32.026 & 0.8231 & \textbf{68.502} & 0.0240 & 13.507 & 0.1719 & \textbf{503.310} & 0.1834  \\
                SRDiff  & 33.506 & 0.8810 & 87.234 & \underline{0.0220} & 13.961 & 0.2103 & 634.908 & \underline{0.1693}   \\
                \hline
			Proposed ASDDPM & 33.684 & 0.8822 & \underline{78.719} & \textbf{0.0217} & 14.170 & 0.2466 & \underline{588.178} & \textbf{0.1677} \\
			\hline
	\end{tabular}}
	\label{tab:SR}
\end{table*}

 \begin{figure*}[!htb]
	\centering
	\includegraphics[width=\linewidth]{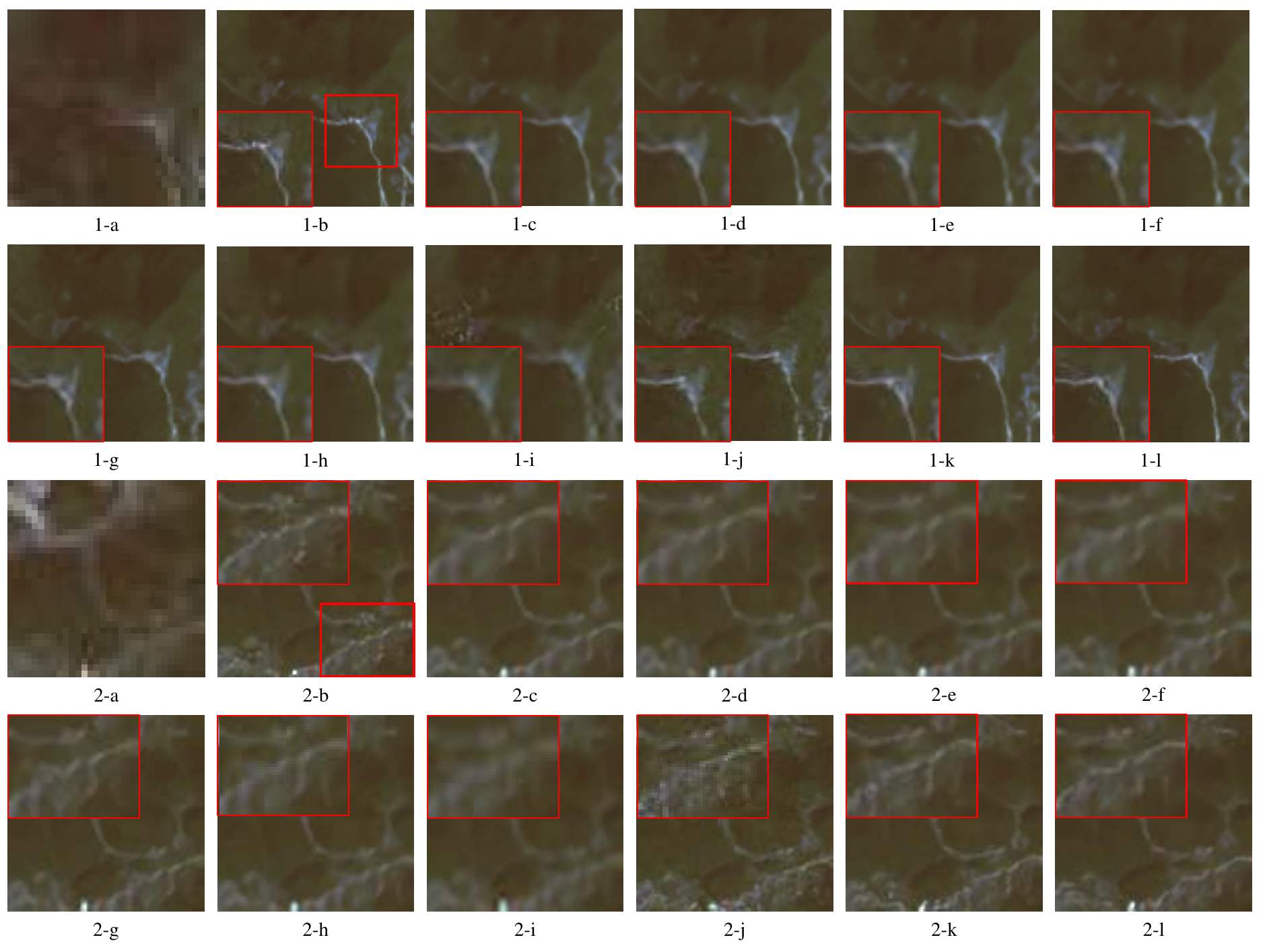}
	\caption{\jl{The visual comparisons of SR results on the OLI2MSI dataset. (a) LR. (b) HR. (c) RDN. (d) NLSN. (e) TranSMS. (f) SRGAN. (g) ESRGAN. (h) Beby-GAN. (i) Dit. (j) EDiffSR. (k) SRDiff. (l) ASDDPM.}}
	\label{fig:OLI2MSI}
\end{figure*}
 \begin{figure*}[!htb]
	\centering
	\includegraphics[width=\linewidth]{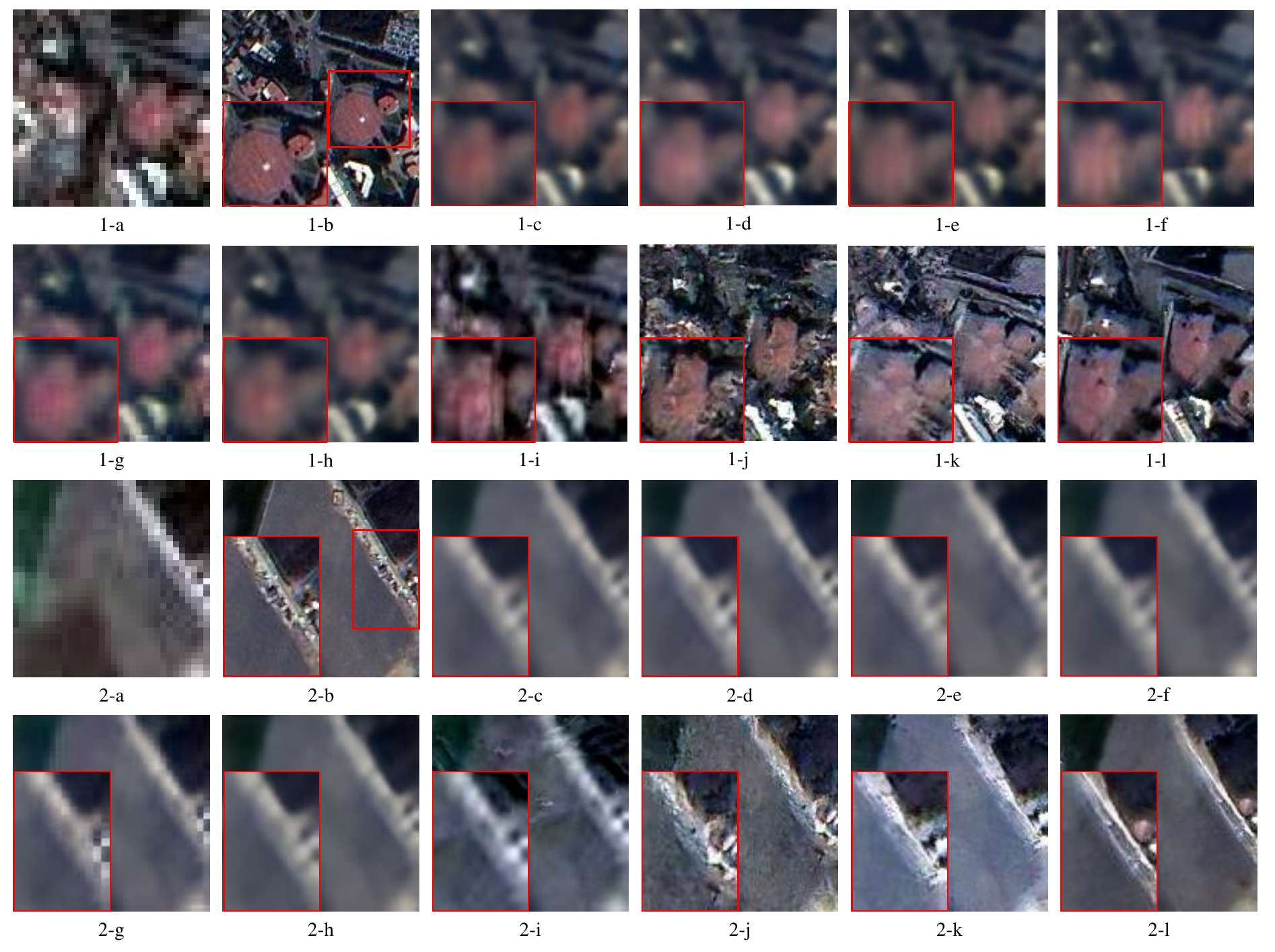}
	\caption{\jl{The visual comparisons of SR results on the OLI2MSI dataset. (a) LR. (b) HR. (c) RDN. (d) NLSN. (e) TranSMS. (f) SRGAN. (g) ESRGAN. (h) Beby-GAN. (i) Dit. (j) EDiffSR. (k) SRDiff. (l) ASDDPM.}}
	\label{fig:Alsat}
\end{figure*}

To evaluate our model, we calculate several performance metrics such as PSNR, SSIM, \jl{Fréchet Inception Distance (FID)}, and Learned Perceptual Image Patch Similarity (LPIPS) \cite{LPIPS} to validate the restored images' quality both at the pixel level and visual perception. More specifically, PSNR measures image quality by calculating the distance between images at the pixel level whereas SSIM focuses on the structural similarity between the two images. \jl{FID is specifically designed for evaluating the results of generative models. It quantifies the Frechet distance between two distributions in the activation space of a pre-trained image classification model, providing a measure of similarity between sets of images.}

Learned Perceptual Image Patch Similarity (LPIPS) is a metric used to measure the perceptual similarity between two images. It is designed to mimic human perception of similarity rather than simply relying on pixel-wise differences. LPIPS has been widely adopted in various computer vision tasks, including image generation, style transfer, and image quality assessment \cite{LPIPS}. It provides a more meaningful visual similarity measure than traditional metrics like PSNR and SSIM, which may not always align with human perception. Specifically, it is calculated based on deep neural networks pre-trained on large datasets containing pairs of images along with human-rated similarity scores. Therefore, it can be used to understand human perceptual judgments of images \cite{srdiff}. 

\subsection{Implementation Details}
Our experiments are implemented with the Pytorch framework and executed on a single NVIDIA GeForce RTX 4090 GPU with $24$~G RAM. The number of channels in the CNP is chosen to be $64$. The kernel size in the CNP is $3$. During the training and testing of DDPM, the number of diffusion steps $T$ for each image is set to $100$ whereas the LR encoder is set as a pre-trained RRDB \cite{srflow}. The noise schedules $\beta_1,\dots,\beta_T$ followed the setting in SRDiff. Specifically, the variable $\beta_t$ is set to  
\begin{equation}
	\beta_t = 1- \frac{\bar{a}_t}{\bar{a}_{t-1}},
\end{equation}
where 
\begin{eqnarray}
\bar{a}_t &=& \frac{f(t)}{f(0)},\\
f(t) &=& \cos\left(\frac{t/T+s}{1+s}\times \frac{\pi}{2}\right),
\end{eqnarray}
with $s = 0.008$ and $\beta_t$ being set to be smaller than 0.99 to prevent singularities at the end of the diffusion process.

Following SRDiff \cite{srdiff}, we choose the $L_1$ loss function to train the model by minimizing the differences between the real and predicted noises. The optimizer is set to Adam with a learning rate of ${10}^{-4}$. 
For the two SR datasets, each pair of LR and HR images in OLI2MSI were randomly cropped into patches of size $32\times32$ and $96\times 96$. For Alsat, we utilized patches of size $32\times32$ and $128\times128$ during the training step. For testing, the images are cropped into patches of the same size as those training patches. For the semantic segmentation task, the original HR images are divided into patches of size $128\times128$ whereas the corresponding LR images are generated by downsampling each HR image to $32\times32$ through the bicubic operation. Finally, we chose $1599$ and $1251$ images as the training and testing set in Vaihingen, respectively. Similarly, $4418$ and $2209$ images in Potsdam are selected as the training and testing sets. 

\subsection{Performance}

For comparison purposes, we evaluate the performance of the proposed ASDDPM against seven state-of-art models, namely three discriminative models RDN \cite{rdn}, NLSN \cite{nlsn}, TranSMS \cite{TranSMS}, four generative models SRGAN \cite{SRGAN}, Beby-GAN \cite{BebyGAN}, ESRGAN \cite{ESRGAN}, Dit \cite{dit}, \jl{EDiffSR \cite{EDiffSR},} and the baseline SRDiff \cite{srdiff}. The test results and visual comparisons of the SISR are shown in Table~\ref{tab:SR} as well as Fig.~\ref{fig:OLI2MSI} and Fig.~\ref{fig:Alsat}. 

Inspection of Table~\ref{tab:SR} shows that the proposed ASDDPM achieved noticeable perceptual quality improvement in LPIPS \jl{and FID}. More specifically, the proposed ASDDPM achieved \jl{LPIPS and FID improvement of $0.0003$ and $9$, $0.0016$ and 46 over the baseline SRDiff on OLI2MSI and Alsat, respectively.} Furthermore, it is shown in Table~\ref{tab:SR} that the proposed ASDDPM achieved balanced performance between sharpness and accuracy in terms of PSNR and SSIM. In sharp contrast, Dit \jl{and EDiffSR} performed poorly in generating detailed and accurate textures on both SR datasets. \jl{Especially, EDiffSR prioritized local information generation, resulting in a significant amount of fake detailed generation. Therefore, it achieved the highest FID score while performing worse in terms of PSNR and SSIM metrics.} In addition, \jl{EDiffSR,} SRDiff and the proposed ASDDPM outperformed other existing methods in terms of LPIPS \jl{and FID} by exploiting the DDPM structure. The GAN-based and discriminative models achieved acceptable performance regarding PSNR and SSIM on the two datasets. This can be attributed to their deliberate utilization of PSNR-oriented loss functions. However, they obtained lower LPIPS \jl{and FID} values due to the worse human perceptual judgments. As shown in Figs.~\ref{fig:OLI2MSI} and \ref{fig:Alsat}, they generated over-smoothed images with significant detail losses. It indicates that higher PSNR and SSIM values cannot ensure better perceptual quality.

\begin{table*}
	\centering
	\caption{The quantitative experimental results (\%) of SR and semantic segmentation on the Vaihingen dataset. The best and second-best are bolded and underlined, respectively.}
	\setlength{\tabcolsep}{4mm}{
		\begin{tabular}{c|cccc|cccc}
			\hline
			    \textbf{Method}&\textbf{PSNR}$\uparrow$&\textbf{SSIM}$\uparrow$&\textbf{FID}$\downarrow$&\textbf{LPIPS}$\downarrow$&\textbf{mF1 (\%) }&\textbf{mIoU (\%) }&\textbf{OA (\%) }\cr
			\hline
                RDN     & \underline{30.683} & \underline{0.8542} & 471.004 & 0.1078 & 86.93 & 77.29 & 88.76  \\
			NLSN    & \textbf{30.823} & \textbf{0.8571} & 478.901 & 0.1051 & 87.06 & 77.49 & 88.91  \\
                TranSMS & 29.884 & 0.8347 & 469.162 & 0.1248 & 85.11 & 74.60 & 87.85  \\
                SRGAN   & 29.531 & 0.8288 & 495.829 & 0.1280 & 83.89 & 72.96 & 87.48  \\
                Beby-GAN& 30.088 & 0.8386 & 337.785 & 0.0879 & 87.18 & 77.71 & 89.25  \\
			ESRGAN  & 30.560 & 0.8494 & 408.982 & 0.0918 & 87.10 & 77.56 & 89.01  \\
                Dit     & 29.234 & 0.8164 & 686.895 & 0.1497 & 84.50 & 73.73 & 87.45  \\
                EDiffSR & 28.502 & 0.7643 & \textbf{83.509} & 0.0496 & 87.54 & 78.21 & 90.10 \\
                SRDiff  & 29.474 & 0.8138 & 112.049 & \underline{0.0432} & \underline{88.24} & \underline{79.42} & \underline{90.39}  \\
			\hline
			Proposed ASDDPM & 29.601 & 0.8186 & \underline{101.545} & \textbf{0.0423} & \textbf{88.41} & \textbf{79.69} & \textbf{90.53}  \\
			\hline
	\end{tabular}}
	\label{tab:SRV}
\end{table*}

\begin{table*}
	\centering
	\caption{The quantitative experimental results (\%) of SR and semantic segmentation on the Potsdam dataset. The best and second-best are bolded and underlined, respectively.}
	\setlength{\tabcolsep}{4mm}{
		\begin{tabular}{c|cccc|cccc}
			\hline
			    \textbf{Method}&\textbf{PSNR}$\uparrow$&\textbf{SSIM}$\uparrow$&\textbf{FID}$\downarrow$&\textbf{LPIPS}$\downarrow$&\textbf{mF1 (\%) }&\textbf{mIoU (\%) }&\textbf{OA (\%) }\cr
			\hline
                RDN     & \underline{33.830} & \underline{0.8790} & 295.599 & 0.0722 & 88.84 & 80.56 & 89.62  \\
			NLSN    & \textbf{33.921} & \textbf{0.8803} & 296.960 & 0.0703 & 88.79 & 80.49 & 89.58  \\
                TranSMS & 33.394 & 0.8710 & 298.616 & 0.0788 & 88.50 & 80.05 & 89.31  \\
                SRGAN   & 32.617 & 0.8564 & 353.783 & 0.0907 & 87.74 & 78.91 & 88.65  \\
                Beby-GAN& 33.177 & 0.8642 & 237.767 & 0.0619 & 88.86 & 80.72 & 89.62  \\
			ESRGAN  & 33.482 & 0.8709 & 260.926 & 0.0665 & 89.03 & 80.82 & 89.69  \\
                Dit     & 31.879 & 0.8364 & 437.622 & 0.1025 & 87.59 & 78.69 & 88.51  \\
                EDiffSR & 31.352 & 0.7930 & \textbf{53.042}  & 0.0491 & 89.75 & 81.83 & 89.92 \\
                SRDiff  & 32.584 & 0.8468 & 87.115 & \underline{0.0410} &\underline{89.92} & \underline{82.22} & \underline{90.52}  \\
			\hline
			Proposed ASDDPM & 32.652 & 0.8461 & \underline{66.190} & \textbf{0.0381} & \textbf{90.29} & \textbf{82.77} & \textbf{90.76}  \\
			\hline
	\end{tabular}}
	\label{tab:SRP}
\end{table*}

Inspection of Fig.~\ref{fig:OLI2MSI} and Fig.~\ref{fig:Alsat} suggests that the SR images restored by the discriminative models including RDN, NLSN, and TranSMS were over-smoothing and blurry. In contrast, the generative models showed improved clarity by predicting and generating information based on the LR images. However, such generative models often suffer from issues such as model collapse, unstable training, and vanishing gradients in practice. In particular, Beby-GAN and ESRGAN showed less satisfactory perceptual quality as compared to discriminative models as these GAN-based models partly compute their content loss at the pixel level. Finally, Fig.~\ref{fig:OLI2MSI} and Fig.~\ref{fig:Alsat} show that the \jl{three DDPM-based models, i.e. EDiffSR, SRDiff and ASDDPM,} generated sharper SR images with greater detail as compared to those GAN-based and discriminative models.

However, since the \jl{DDPM-based models, such as SRDiff and EDiffSR,} generates SR details based on noise prediction learned from the training set, the errors incurred in each diffusion time step are amplified through iterations, which entails large discrepancies between the input and SR images. In contrast, the proposed ASDDPM could provide more detailed and reliable texture information under these circumstances as evidenced in Fig.~\ref{fig:Alsat}. \jl{The feature fusion between the two decoders took full advantage of the feature from U-Net encoder by preserving the high- and low-frequency semantic information of LR images, which resulted in better perceptual quality.} For instance, for the ``road" in the second row, the SR image generated by the proposed ASDDPM closely resembled the HR image with clear boundaries whereas the SRDiff-generated image exhibited patterns different from the HR image.

 \begin{figure*}[!htb]
	\centering
	\includegraphics[width=\linewidth]{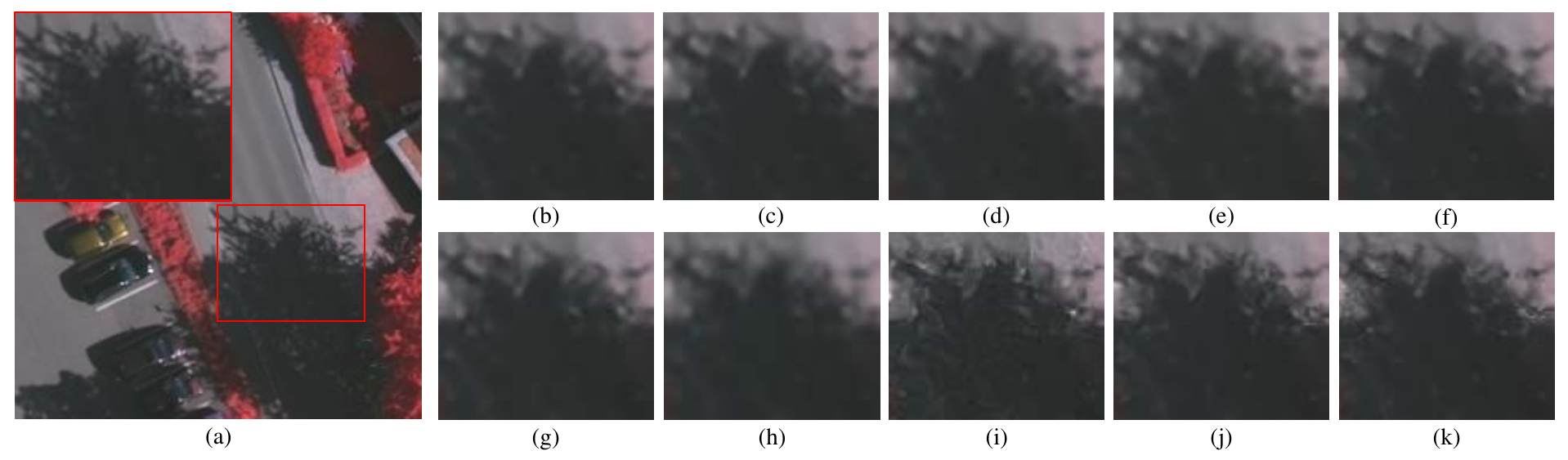}
	\caption{\jl{The visual comparisons of SR results on the Vaihingen dataset.  (a) HR. (b) RDN. (c) NLSN. (d) TranSMS. (e) SRGAN. (f) ESRGAN. (g) Beby-GAN. (h) Dit. (i) EDiffSR. (j) SRDiff. (k) ASDDPM.}}
	\label{fig:SRV}
\end{figure*}

 \begin{figure*}[!htb]
	\centering
	\includegraphics[width=\linewidth]{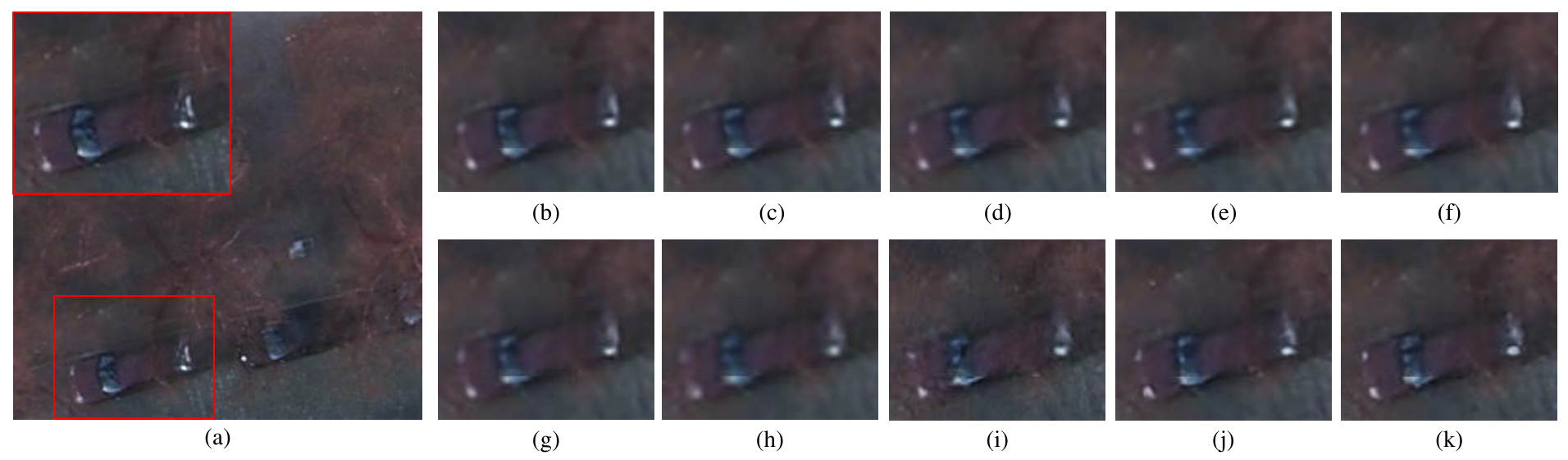}
	\caption{\jl{The visual comparisons of SR results on the Potsdam dataset. (a) HR. (b) RDN. (c) NLSN. (d) TranSMS. (e) SRGAN. (f) ESRGAN. (g) Beby-GAN. (h) Dit. (i) EDiffSR. (j) SRDiff. (k) ASDDPM.}}
	\label{fig:SRP}
\end{figure*}

\begin{figure*}[h]
	\centering
	\includegraphics[width=0.7\linewidth]{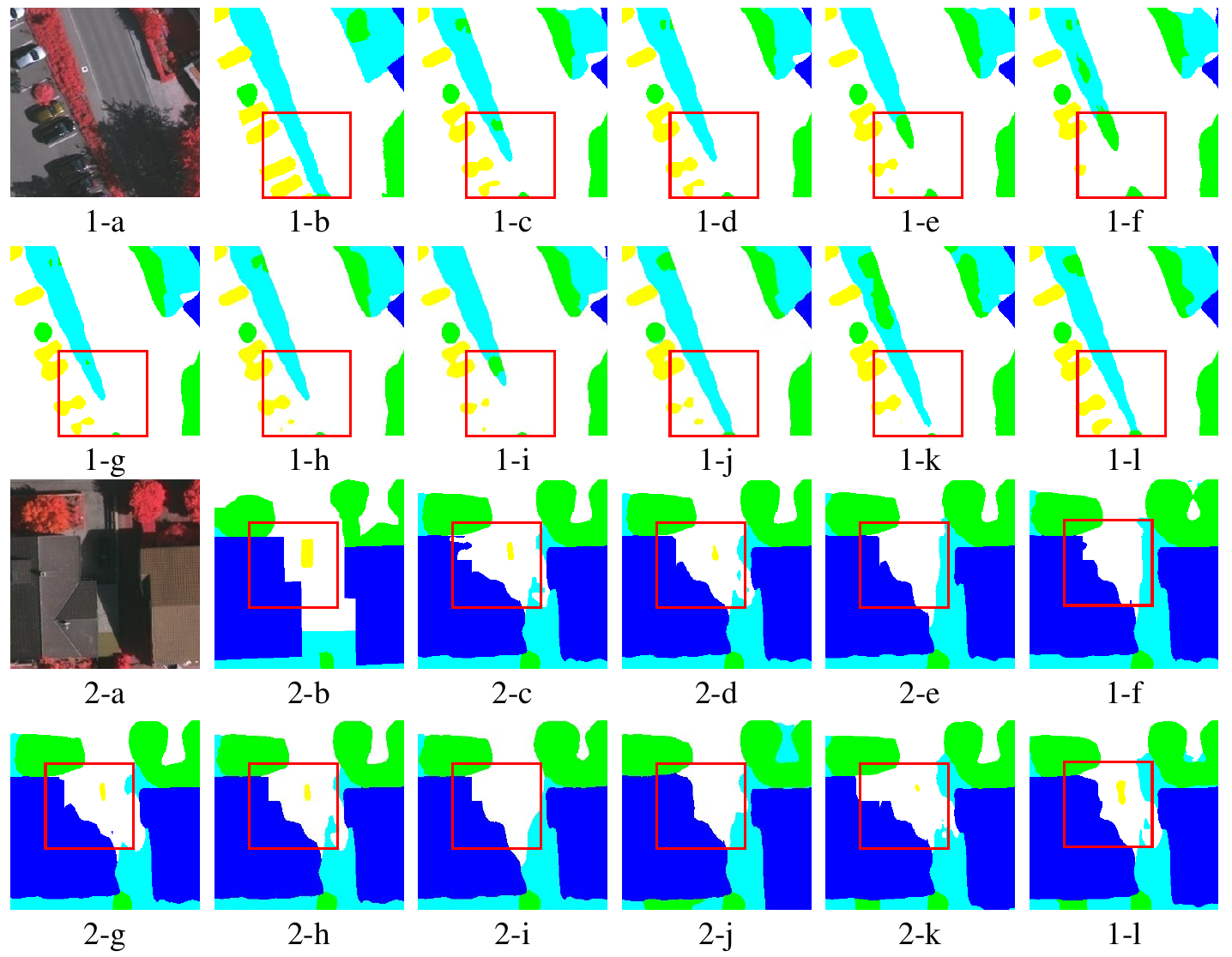}
	\caption{\jl{The semantic segmentation comparisons on the Vaihingen dataset. (a) Real image. (b) Ground truth. (c) RDN. (d) NLSN. (e) TranSMS. (f) SRGAN. (g) ESRGAN. (h) Beby-GAN. (i) Dit. (j) EDiffSR. (k) SRDiff. (l) ASDDPM.} Each color represents the following. Blue: Building. Green: Tree. Cyan: Low Veg. Yellow: Car. White: Imp Suf. Red: clutter/background.}
	\label{fig:Vaihingen}
\end{figure*}

 \begin{figure*}[h]
	\centering
	\includegraphics[width=0.7\linewidth]{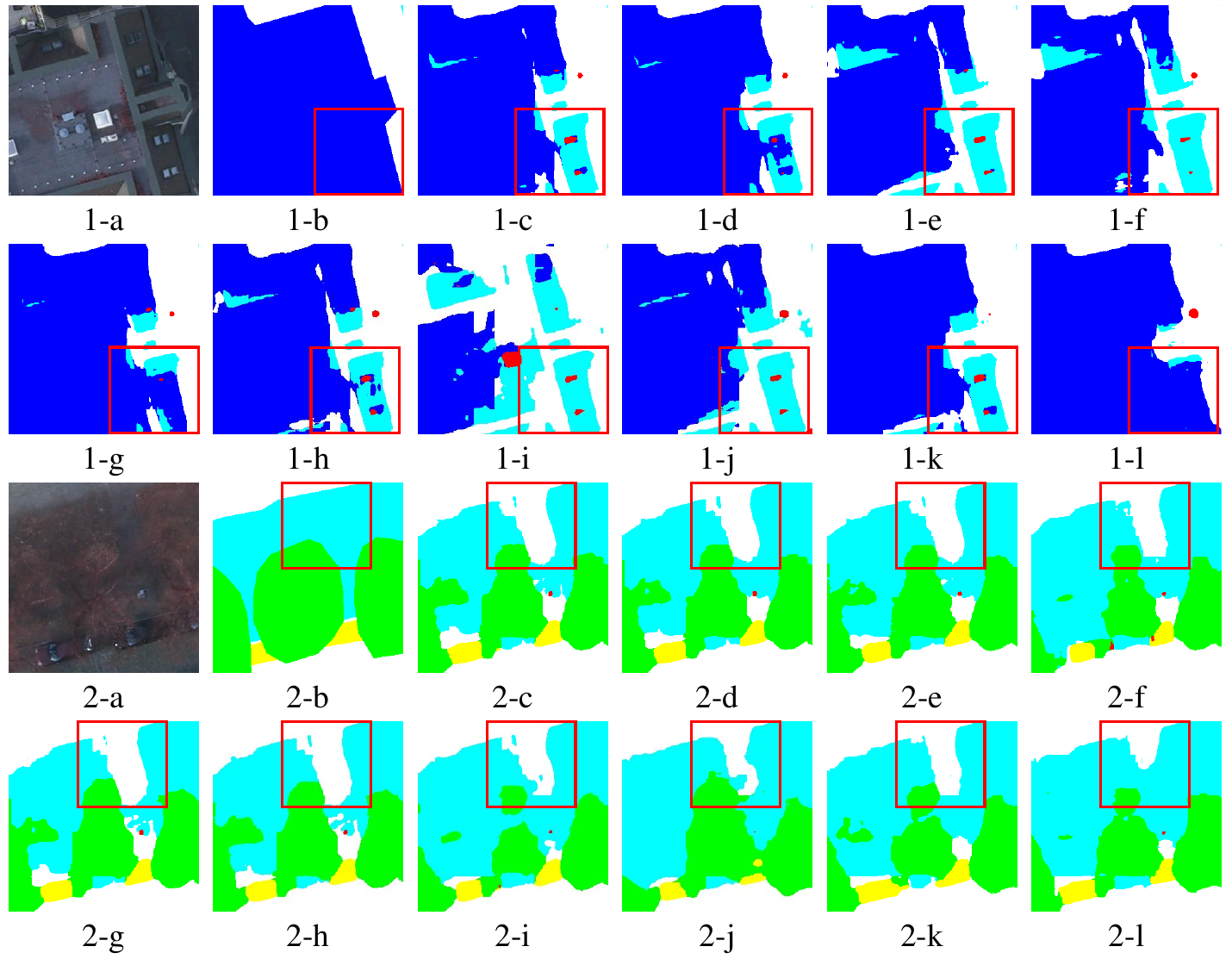}
	\caption{\jl{The semantic segmentation comparisons on the Potsdam dataset. (a) Real image. (b) Ground truth. (c) RDN. (d) NLSN. (e) TranSMS. (f) SRGAN. (g) ESRGAN. (h) Beby-GAN. (i) Dit. (j) EDiffSR. (k) SRDiff. (l) ASDDPM.} Each color represents the following. Blue: Building. Green: Tree. Cyan: Low Veg. Yellow: Car. White: Imp Suf. Red: clutter/background.}
	\label{fig:Potsdam}
\end{figure*}

\subsection{Performance on Semantic Segmentation Task}
Two HR segmentation datasets, namely Vaihingen and Potsdam, were employed to validate the performance of the synthesized SR images in semantic segmentation tasks. More specifically, the HR images in these two datasets are first resized with the bicubic operation into $1/4$ of their original size before different SISR models are applied to generate the corresponding SR images. After assessing the quality of the SR images using PSNR, SSIM, LPIPS, \jl{and FID}, we feed the SR images into the same pre-trained U-Netformer \cite{unetformer} to evaluate the corresponding segmentation performance in terms of the overall accuracy (OA), the mean intersection over union (mIoU), and the mean F1 score (mF1).

Specifically, mF1 and mIoU are computed based on five ground classes, namely ``roads", ``buildings", ``low vegetation", ``trees", and ``cars", while the evaluation of OA includes one more class named ``clutter". The definition of the OA is given by:
\begin{equation}
	OA = \frac{TP}{TP+FP+TN+FN},
\end{equation}
where TP, FP, TN, and FN represent the number of true positives, false positives, true negatives and false negatives, respectively. Moreover, F1 and IOU are computed based on each category according to the following formula:
\begin{eqnarray}
	F1 &=& 2 \times \frac{Q \times R}{Q + R},\\
	IoU& =& \frac{TP}{TP+FN+FP},
\end{eqnarray}
where $Q$ and $R$ are given by:
\begin{eqnarray}
	Q &=&  \frac{TP}{TP+FP},\\
	R &=&  \frac{TP}{TP+FN}.
\end{eqnarray}
Finally, mF1 and mIoU are derived from the mean of F1 and IoU in each category according to the formula above.

\begin{table}[h]
	\centering
	\caption{The quantitative experimental results (\%) for five main classes on the Vaihingen dataset.}
	\setlength{\tabcolsep}{2mm}{
		\begin{tabular}{c|ccccc}
			\hline
			\multirow{2}{1cm}{\textbf{Method}}&\multicolumn{5}{c}{\textbf{IoU (\%) }}\cr
		    &\textbf{roads}&\textbf{buildings}&\textbf{low veg}&\textbf{trees}&\textbf{cars}\cr
			\hline
                RDN     & 81.64 & 87.97 & 63.91 & 80.77 & 72.17 \\
			NLSN    & 81.84 & 88.26 & 64.10 & 80.97 & 72.27 \\
                TranSMS & 80.48 & 86.41 & 62.18 & 79.59 & 64.36 \\
                SRGAN   & 79.74 & 85.74 & 61.72 & 79.60 & 58.00 \\
                Beby-GAN& 82.46 & 89.23 & 64.52 & 81.07 & 71.26 \\
			ESRGAN  & 82.13 & 88.66 & 64.18 & 80.83 & 72.02 \\
                Dit     & 79.64 & 85.90 & 60.90 & 79.35 & 62.86 \\
                EDiffR  & 84.37 & 90.99 & 63.48 & 81.39 & 70.82 \\
                SRDiff  & 84.24 & 91.35 & 65.67 & 82.52 & 73.05 \\
			\hline
			Proposed ASDDPM & \textbf{84.86} & \textbf{91.75} & \textbf{65.74} & \textbf{82.52} & \textbf{73.57}  \\
			\hline
	\end{tabular}}
	\label{tab:Vdetail}
\end{table}

\begin{table}[h]
	\centering
	\caption{The quantitative experimental results (\%) for five main classes on the Potsdam dataset.}
	\setlength{\tabcolsep}{2mm}{
		\begin{tabular}{c|ccccc}
			\hline
			\multirow{2}{1cm}{\textbf{Method}}&\multicolumn{5}{c}{\textbf{IoU (\%) }}\cr
		    &\textbf{roads}&\textbf{buildings}&\textbf{low veg}&\textbf{trees}&\textbf{cars}\cr
			\hline
                RDN     & 82.97 & 92.24 & 70.79 & 65.87 & 90.88 \\
			NLSN    & 82.99 & 92.26 & 70.59 & 65.52 & 91.05 \\
                TranSMS & 82.65 & 91.91 & 69.93 & 64.85 & 90.89 \\
                SRGAN   & 81.74 & 91.34 & 68.50 & 62.68 & 90.28 \\
                Beby-GAN& 82.87 & 92.05 & 71.05 & 66.47 & 91.14 \\
			ESRGAN  & 82.82 & 92.14 & 71.38 & 66.68 & 91.03 \\
                Dit     & 81.51 & 91.04 & 68.69 & 61.95 & 90.25 \\
                EDiffSR & 83.08 & 92.01 & 70.78 & 71.44 & 91.87 \\
                SRDiff  & 84.85 & 92.92 & 71.38 & 70.12 & 91.81 \\
			\hline
			Proposed ASDDPM & \textbf{84.93} & \textbf{92.97} & \textbf{72.31} & \textbf{71.48} &  \textbf{92.14}  \\
			\hline
	\end{tabular}}
	\label{tab:Pdetail}
\end{table}

Table~\ref{tab:SRV} and Table~\ref{tab:SRP} show the SR quality and the resulting segmentation performance based on the Vaihingen and Potsdam datasets, respectively. In both tables, the our proposed ASDDPM, showed comparable PSNR and SSIM performance as compared to the discriminative and GAN-based models. However, ASDDPM achieved substantially improved LPIPS \jl{and FID} performance as well as segmentation performance in terms of mF1, mIoU, and OA values. As shown in Table~\ref{tab:SRV} derived from the Vaihingen dataset, ASDDPM achieved performance improvements of about $1.3\%$ in mF1, $2.1\%$ in mIoU, and $1.5\%$ in OA respectively, as compared to the best performing GAN-based model, i.e. ESRGAN. Similarly, the proposed ASDDPM outperformed the best-performing discriminative models, i.e. NLSN by $1.4\%$ in mF1, $2.2\%$ in mIoU, and $1.6\%$ in OA, respectively. Furthermore, ASDDPM outperformed SRDiff by a large margin of $0.17\%$, $0.27\%$, and $0.14\%$ in terms of mF1, mIoU, and OA, respectively. Similar observations can be obtained from Table~\ref{tab:SRP} derived from the Potsdam dataset. Finally, Fig.~\ref{fig:Vaihingen} and Fig.~\ref{fig:Potsdam} present visualization of the segmentation results achieved with various models. It is evidenced from Fig.~\ref{fig:Vaihingen} and Fig.~\ref{fig:Potsdam} that the proposed ASDDPM provided more accurate and reliable predictions in shape and outline with fewer erroneous class assignments.

Table~\ref{tab:Vdetail} and Table~\ref{tab:Pdetail} summarize the IoU performance of various models on each of the five categories. Inspection of Table~\ref{tab:Vdetail} and Table~\ref{tab:Pdetail} suggests that the DDPM-based models outperformed the discriminative and GAN-based models across all classes. In particular, the proposed ASDDPM achieved the best performance among all models. For instance, in Table~\ref{tab:Vdetail}, ASDDPM outperformed SRDiff by $0.6\%$, $0.4\%$, $0.1\%$, and $0.5\%$ in the roads, buildings, low vegetation, and cars. A similar observation was obtained for the Potsdam dataset as shown in Table~\ref{tab:Pdetail} in which ASDDPM achieved improved segmentation performance of $0.1\%$, $0.05\%$, $1\%$, $1.3\%$, and $0.3\%$ in terms of ``roads", ``buildings", ``low vegetation", ``trees", and ``cars". This observation can be explained by the fact that HR images with elevated LPIPS \jl{and FID} values generated by the ASDDPM exhibit superior perceptual quality. The presence of accurate and rich detailed textures proves beneficial for distinguishing object types, which further supports our argument that pixel-level performance metrics such as PSNR and SSIM are inadequate for evaluating SR performance.

\begin{table}[h]
	\centering
	\caption{Ablation study on the OLI2MSI for U-Net decoder, ADTD and FI module.}
	\setlength{\tabcolsep}{1.3mm}{
		\begin{tabular}{ccccccc}
			\hline		
                \centering\textbf{U-Net decoder}&\textbf{ADTD}&\centering\textbf{FI}& \textbf{PSNR} &\textbf{SSIM}&\textbf{LPIPS}&\textbf{FID} \\
                \hline
                &\checkmark&& 32.944 & 0.8654 & 0.0237 & 84.329\\
                \checkmark&&& 33.506 & 0.8810 & 0.0220 & 87.234\\
			\checkmark&\checkmark&& 33.679 & 0.8814 & 0.0218 & 83.707\\
			\checkmark&\checkmark&\checkmark&\textbf{33.684} & \textbf{0.8822}&\textbf{0.0217}& \textbf{78.719}\\
		\hline
	\end{tabular}}
	\label{tab:Ablation}
\end{table}

\begin{table}[h]
	\centering
	\caption{\jl{Ablation study on the OLI2MSI for time step in sampling step.}}
	\setlength{\tabcolsep}{1.3mm}{
		\begin{tabular}{cccccc}
			\hline		
                \centering\textbf{Time Step}&\textbf{PSNR} &\textbf{SSIM}&\textbf{LPIPS}&\textbf{FID} \\
                \hline
                
			10 & 13.003 & 0.0302 & 0.8285 & 5976.626\\
                50 & 20.033 & 0.4307 & 0.1629 & 2490.093\\
			100&\textbf{33.684} & \textbf{0.8822}&\textbf{0.0217}& \textbf{78.719}\\
                200& 28.365 & 0.6508 & 0.0640 & 508.893\\
		\hline
	\end{tabular}}
	\label{tab:step}
\end{table}
\subsection{Ablation Study}
Table~\ref{tab:Ablation} presents the ablation experiment results derived from OLI2MSI to evaluate the efficiency of the U-Net decoder, the ADTD, and the FI modules. The first two rows of experimental results were obtained with the CNP equipped with a single decoder, either the ADTD or the U-Net decoder. Compared with the CNP with a single decoder, the CNP with a dual-decoder structure achieved improved image reconstruction performance as indicated in the third row. For instance, the CNP with a dual-decoder structure outperformed that with a single ADTD by $0.735$ dB, $0.016$, $0.002$, \jl{and $0.6$} in terms of PSNR, SSIM, LPIPS, \jl{and FID}, respectively. Finally, the addition of the FI modules helped further improve the performance by $0.005$ dB, $0.001$, $0.0001$, \jl{and $5$} in terms of PSNR, SSIM, LPIPS, \jl{and FID}, respectively. The ablation experiments confirmed the effectiveness of the proposed dual-decoder structure and the FI modules in the proposed ASDDPM.

\jl{We conducted experiments to test the influence of the time step in the sampling process. As shown in Table~\ref{tab:step}, low time step, such as 10 and 50, resulted in significantly worse performance degradation, while quite high time step, such as 200, still led to performance degradation. In summary, our model appeared superior performance with time step 100.}

\subsection{Noise Prediction Performance}


\begin{figure}
    \centering
    \subfloat[]{
	\includegraphics[scale=0.3]{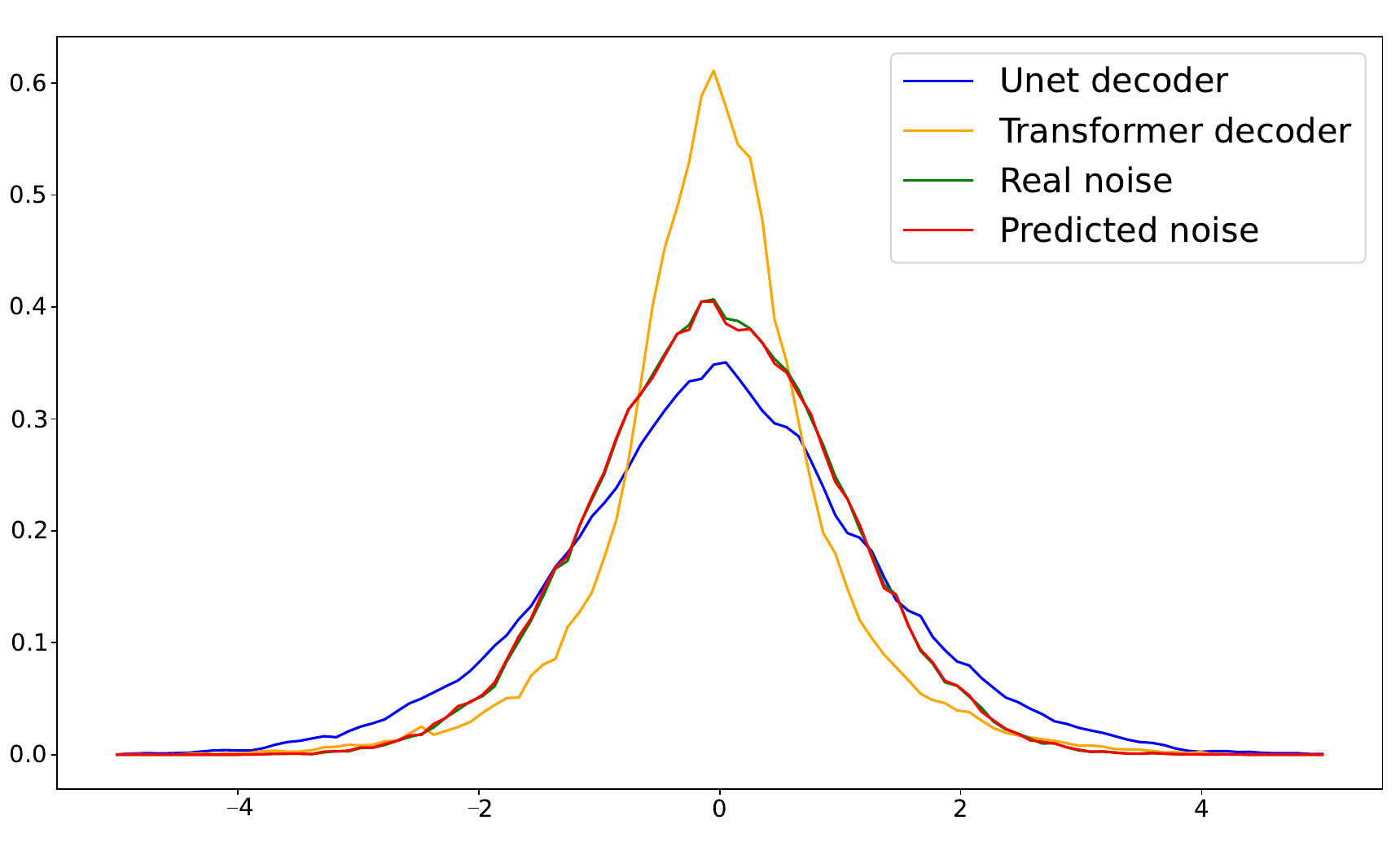}
    }
    
    \subfloat[]{
	\includegraphics[scale=0.3]{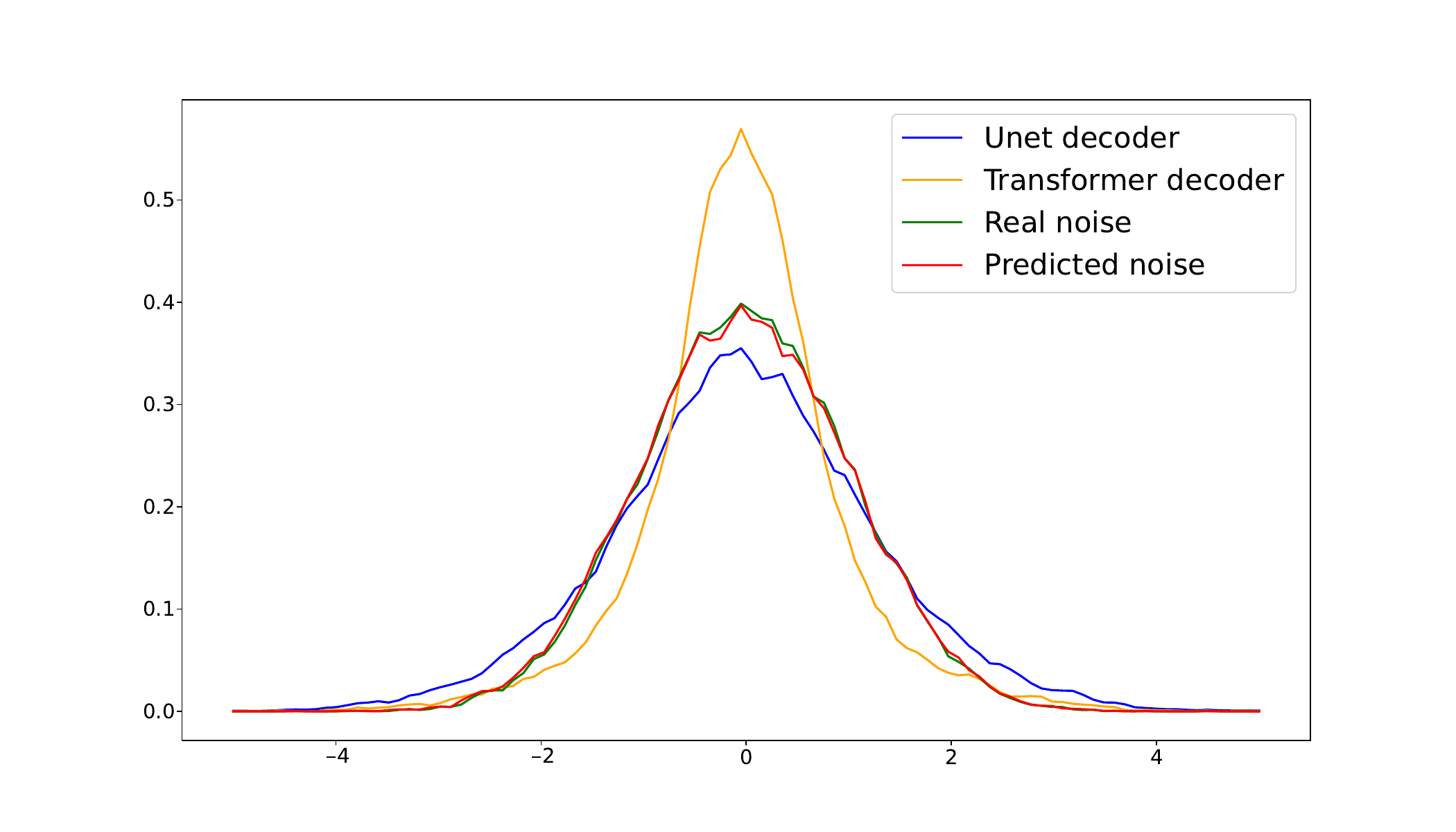}
    }
    \caption{Illustration of the noise prediction performance of the ADTD and U-Net decoder on datasets of (a) OLI2MSI and (b) Alsat.}
    \label{fig:decoder}
\end{figure}

Fig.~\ref{fig:decoder} depicts examples of the cumulative distribution functions (CDF) of the ADTD output, the U-Net decoder output, real noise, and the final predicted noise. In this experiment, the FI was removed to highlight the characteristics of the U-Net decoder and ADTD. While all CDFs followed approximately the Gaussian distribution, the CDF of the predicted noise generated by the proposed ASDDPM was almost identical to that of the real noise. Furthermore, the CDF of the U-Net decoder output showed a larger variance as compared with that of the ADTD output. This is because the U-Net decoder usually provides more {\em specific} information with prominent values, whereas the ADTD generates more general information with more homogeneous characteristics. The utilization of the two decoders resulted in better performance in terms of both the pixel-level metrics and the visual perception, with reduced misalignment and less fake textures.

\begin{table}[h]
	\centering
	\caption{Computational complexity analysis of all methods.}
	\setlength{\tabcolsep}{1.5mm}{
		\begin{tabular}{cccccc}
			\hline
			\multirow{2}{1cm}{\textbf{Method}}&\textbf{Models}&\textbf{Complexity}&\textbf{Memory}&\textbf{Parameters}&\textbf{Speed}\\
     	&&\textbf{(GFLOPs)}&\textbf{(MB)}&\textbf{(M)}&\textbf{(FPS)} \\
			\hline
			&RDN&365.56&5231&22.31&0.786\\
                Discriminative&TranSMS&135.12&8413&4.93&0.689\\
			&NLSN&733.69&6877&44.75&0.772\\
			\hline
			\multirow{3}{1.5cm}{GAN-based}&SRGAN&14.69&1653&0.73&0.800 \\
                &Beby-GAN&399.71&10318&23.17&0.903 \\
			&ESRGAN&9.97&1537&0.62& 0.825\\

			\hline
                 \multirow{4}{1.6cm}{DDPM-based}&Dit&225.16&6356&33.13&0.006\\
                 &EDiffSR&174.61&1954&30.39&0.077\\
                 &SRDiff&186.08&2842&11.66&0.012\\
			&ASDDPM&262.40&3822&13.77&0.008\\
			\hline
	\end{tabular}}
	\label{tab:models}
\end{table}
\subsection{Computational Complexity Analysis}
Finally, we compare the computational complexity of all methods in terms of model complexity, memory usage, the number of parameters, and inference speed. Specifically, the model complexity was evaluated through Giga Floating-point Operations Per Second (GFLOPs), where $1~GFLOPs =10^{9}~FLOPs$. The GPU memory, measured in megabytes (MB), is provided alongside the number of parameters, measured in millions (M), and the inference speed, measured in frames per second (FPS).

All models were divided into three categories, namely the discriminative, the GAN-based, and the DDPM-based models. The results presented in Table~\ref{tab:models} indicate that DDPM-based models exhibited higher computational complexity, leading to considerably slower inference speeds compared to other models. Notably, the proposed ASDDPM performed comparably to other DDPM-based models but with lower computational costs and higher inference speed as compared to Dit.


\section{Conclusion} \label{sec:conclusion}

This work has proposed a novel model called ASDDPM for SR image reconstruction by exploiting a DDPM with a dual-decoder architecture explicitly designed for remote sensing applications. \jl{The proposed ASDDPM bridges the semantic gap between the encoder and decoder by incorporating an enhanced decoder, ADTD. It introduces low-frequency semantic information to control the detailed texture generation, enabling the CNP to produce more accurate noise distributions. As a result, it overcomes the low accuracy and reliability issues encountered by DDPM as well as generates sharper details and texture.} Extensive experiments conducted on four satellite remote sensing datasets, namely Alsat, OLI2MSI, Vaihingen, and Potsdam, have confirmed the effectiveness and generalization capability of the proposed ASDDPM. The main limitation of the proposed approach lies in the relatively higher computational cost because the denoising process is implemented on the image space. Therefore, future studies will focus on the training optimization process of ASSDDPM for denoising in the latent space instead of in the image space. This change has the potential of reducing computational complexity without sacrificing flexibility or quality.

\small
\bibliographystyle{IEEEtranN}
\bibliography{references}

\end{document}